# Spectroscopic Evidence of Low Energy Gaps Persisting Towards 120 Kelvin in Surface-Doped *p*-Terphenyl Crystals

Haoxiang Li[1†*], Xiaoqing Zhou[1†*], Stephen Parham[1], Thomas Nummy[1], Justin Griffith[1], Kyle Gordon[1], Eric L. Chronister[2] and Daniel S. Dessau[1,3*]

[1]*Department of Physics, University of Colorado at Boulder, Boulder, CO 80309, USA*
[2]*Department of Chemistry, University of California, Riverside, CA 92521, USA*
[3]*Center for Experiments on Quantum Materials, University of Colorado at Boulder, Boulder, CO 80309, USA*

† These authors contributed equally to this work

* Correspondence to: Haoxiang.Li@Colorado.edu

　　　　　　　　　　Xiaoqing.Zhou@Colorado.edu

　　　　　　　　　　Dessau@Colorado.edu

**The possibility of high temperature superconductivity in organic compounds has been discussed since the pioneering work of Little in 1964 [1], with unsatisfactory progress until the recent report of a weak Meissner shielding effect at 120 Kelvin in potassium-doped *para*-terphenyl samples [2]. To date however, no other signals of the superconductivity have been shown, including the zero-resistance state or evidence for the formation of the Cooper pairs that are inherent to the superconducting state. Here, using high-resolution photoemission spectroscopy on potassium surface-doped *para*-terphenyl crystals, we uncover low energy gaps that persist to approximately 120 K. Among a few potential origins for these gaps, we argue that the onset of electron pairing within molecules is most likely. And while pairing gaps are a prerequisite for high temperature superconductivity they do not guarantee it. Rather, the development of long-range phase coherence between the paired states on the molecules is necessary, requiring good wavefunction overlap between molecular states – something that is in general difficult for such weakly overlapping molecules.**

*Para*-terphenyl is a simple organic molecule composed of three benzene rings arranged end-to-end, as illustrated in Fig. 1a, and it is available commercially at a modest price. These molecules can be packed together in single crystalline form (Fig. 1c), in which case the molecules arrange themselves in a unidirectional stacking as shown in Fig. 1b. The experiments that reported the weak Meissner shielding effect above 120K [2, 3] were prepared from non-crystalline *p*-terphenyl powders, annealed with potassium in an evacuated tube.

The Meissner effect is just one signature of superconductivity, so it is in need of confirmation from other techniques such as transport or spectroscopy, with the latter also able to give critical information needed to understand the origin of the possible superconductive pairing. The Meissner effect signal in these papers was also extremely weak, with a volume fraction about 0.04%. This implies that only a tiny fraction of the end products became superconducting, or that any superconducting pairs were constrained to very short length scales. In this paper, we present a photoemission study on pristine *p*-terphenyl single crystals (Fig. 1c) with controlled in-situ potassium metal (K) evaporation in ultra-high vacuum, aiming to directly detect the presence of the Cooper pairs that are at the heart of all known superconductors.

Fig. 1d shows a schematic of the experiment. The crystals were initially annealed in ultrahigh vacuum at 100C, which due to the high vapor pressure of *p*-terphenyl will sublimate off any dirty exterior layers. Sub monolayer coverages of K were then consecutively dosed onto the clean surface at T=300 K, with the surfaces monitored by x-ray core level spectroscopy (XPS) as well as by high-resolution photoemission of the near-Fermi level features. Fig. 1e shows spectra for a variety of consecutive doses. There are at least 4 peaks in the spectra at binding energies (energy below $E_F$) near 5 eV, 7 eV, 9 eV and 13.5 eV respectively, corresponding to various peaks in the valence band/occupied molecular orbitals. For the pristine compound, there is vanishingly small spectral weight for the first 2 eV below the chemical potential, consistent with the optical gaps that are of order 3-4 eV [4]. With consecutive K surface dosings, a potassium 3p core level develops at the binding energy around 18 eV, indicating that potassium is incorporated onto/into the surface. Even in the presence of K-dosing, the original valence peak features remain robust. This indicates that the potassium doping is perturbative in nature, only minimally modifying the large-scale electronic structure of *p*-terphenyl. On the other hand, the peak positions were monotonically shifted away from $E_F$, indicating a change in chemical potential and the spectral weight in the vicinity of the chemical potential grows (not visible in the wide scale scan of Fig. 1e). This is consistent with the idea that potassium donates extra electrons to the lowest energy conduction bands.

To date, minimal angle-dependent changes have been observed, which is presumably due to two possibilities. A) The very weak dispersion expected in organic crystals in which the constituent components are far separated with weak orbital overlap. A flat dispersion [5] combined with strong electronic correlations can cause a strong broadening effect, and heavily smear out the momentum-dependent features. B) Possible disorder of the underlying crystal lattice or K overlayers, which were not annealed after the K deposition can provide extra scattering of the electron, which would smear the momentum dependent features. For this reason, the present spectra are not labeled by momentum-space positions, and should be viewed as representing the average effect across the Brillouin zone.

With sufficient K-dosing, very weak metallic spectral weight appeared near the chemical potential and the material became much more conductive, as also evidenced by the lack of sample charging at low temperatures (see Fig. S1). This result has similarities to some other doped aromatic organic compounds such as picene and coronene, in which minimal electronic weight was found at the Fermi level [6,7], and is different from doped $C_{60}$ that does show strong spectral weight at $E_F$ [8,9]. A more detailed discussion about the metallic but weak spectral weight at $E_F$ is contained in the supplementary information S1 and Fig. S2.

Fig. 2a shows the very low energy regime for sample #3 as a function of temperature between 10 K and 200 K (see Figs. S3 and S5 for some data on samples 1 and 2). The leading edge of the 10 K spectrum is pulled away from the chemical potential, as also evidenced by an overlay of this spectrum with that from a metallic gold film measured under identical conditions right after the measurement of the doped *p*-terphenyl (see Fig. S4). An alternative view of the same spectra is presented in Fig. 2b, which shows the data of Fig. 2a symmetrized about $E_F$, which has been developed as a powerful way to remove the effect of the Fermi function and better visualize the presence of any low energy gaps [10]. Here we see that there is a strong suppression of low energy spectral weight (a gap or pseudogap) that gradually disappears as the temperature is raised. Fig. 2c shows the spectral weight lost at $E_F$ (integrated over ± 3 meV) that is removed by the low energy gap, normalized to a maximum effect of 1 at our lowest temperature. It is seen that the low energy spectral weight is fully recovered at a temperature near 120 K, above which there are minimal changes. This temperature is within error the same as the observed onset of the weak Meissner effect in ref [2], raising the possibility that the gaps we observe are related to the Meissner effect, that is, to the formation of Cooper pairs.

Other possible origins for the observed gaps or pseudogaps, such as due to a charge density wave (CDW), spin density wave (SDW), Coulomb gap, or polaronic effects are less plausible than that of a pairing gap, because a) no evidence for a CDW or SDW so far exists in these compounds, b) a Coulomb gap or polaronic gap would usually have a much "softer" or slowly varying character as a function of energy, c) none of the others would be expected to have the temperature dependence shown in Fig. 2c, while a pairing gap would, especially considering the onset temperature of the Meissner effect. Such a temperature evolution of the spectral weight lost was shown as a signature behavior in the pairing gap of cuprate high $T_C$ superconductors (see Fig. 3 in ref. [11]). A more detailed discussion of the origin of the spectral gap is contained in supplementary information S2. Regardless of this, it is noted that the Meissner effect remains weak and that no clear evidence for zero electrical resistance has yet been reported. Possible reasons for this will be discussed near the end of the paper.

The gap data of Fig. 2b clearly shows the gap filling behavior with temperature, similar to the cuprate high temperature superconductors [12,13,14]. Such a "filling-in" behavior as well as weak or absent

"coherence peaks" at the gap edge is most commonly and simply modeled with the Dynes model for the superconducting density of states $N_{SC}(E,T)$ [15]:

$$N_{SC}(E,T) = N_N(E)\, Re\left(\frac{E - i\Gamma(T)}{\sqrt{(E-i\Gamma(T))^2 - \Delta(T)^2}}\right), \tag{1}$$

where $N_N(E)$ is the normal state density of states and $\Gamma(T)$ represents a scattering rate or pair-breaking effect that competes with the superconducting gap $\Delta(T)$. The dotted lines in Fig. 2b show fits to the experimental data using this equation convolved with the measured experimental resolution function, and with $N_N(E)=a+bE$, i.e. a linearly varying density of states. The parameters extracted from these fits are shown in Fig. 2d. The gap has a low temperature magnitude of $\Delta(0)\sim 12$ meV and is roughly constant as the temperature is increased to 60K or above. The fits also show that $\Gamma$, which breaks the pairs, starts small and rises rapidly with temperature, which is unexpected for a conventional BCS superconductor but is a well-known characteristic for cuprate high temperature superconductors (see Fig. S7 and in refs. [12,13,14]). With $\Delta$ considered as a pairing gap, our fits show that $\Gamma$ becomes larger than $\Delta$ at approximately 60 K. Above this temperature the rate at which pairs are broken will be faster than the rate at which they are created, so we expect that the formation of long range phase coherence above 60 K would be especially difficult. Our results above 60K are therefore also strongly reminiscent of the pseudogap (pre-pairing) state in the cuprates, which are widely (but not universally) discussed as a state with preformed Cooper pairs that have not yet condensed into the phase-coherent SC state [11,13,16].

The success of the simple Dynes model fitting, as well as the general resemblance of our observed phenomenology to that of cuprate superconductors, enhances the likelihood of high temperature superconductivity in this class of materials, and indicates that the host of the superconductivity should be the K-doped or K-intercalated *p*-terphenyl itself. In contrast to the original report[2, 17] in which the end product is arguably a mixture of different components, in our experiment it is highly unlikely that chemical reactions substantially modified the material phase. Our results also suggest two main possibilities for the weak Meissner effect and the absence of the zero-resistance state in measurements to date: 1) This is potentially due to a very weak overlap between various grains in the existing bulk doped (polycrystalline) compounds, which would have to be Josephson coupled to enable superconductivity, therefore leaving a weakened Meissner effect with all other superconducting properties weakened as well [18]. 2) Alternatively, the weak signals might be due to the intrinsically weak electronic overlap between individual terphenyl molecules, which could host local pairing similar to that in a Bose-Einstein Condensation (BEC) picture [19], but have a difficult time fully condensing into a coherent superconducting state with high phase rigidity. This is partly evidenced by the large pair-breaking Γ term shown in Fig. 2d, by the weak Meissner signal, and by the lack of zero resistance so far observed (see a more detailed discussion in supplementary information S1).

The mechanism of the likely pairing in these materials is also potentially quite different from other known superconductors, not just because of the high $T_c$'s but also because of the unusual structure and chemistry of organic molecular solids. Little's original proposal suggested that a fully electronic (non-phononic) mechanism may be possible [1] and other proposals for organic superconductors including Resonating Valence Bond (RVB) physics [20] as well as bipolaronic pairing mechanisms [21] have been discussed in the context of organic superconductivity. However, since the proposed bipolaron energy scale from Raman spectroscopy is of the order of 180 meV in these materials [2], it is not obvious if we can connect bipolaron physics to the much lower energy scale gaps that are observes here. The present findings therefore potentially open new and exciting venues into the most fundamental aspects of superconductivity as well.

**Note:** After the submission of this manuscript other groups have made progress on this topic. A Meissner effect above 120 K with 20 times stronger signal was reproduced on high pressure synthesized *K*-doped *p*-terphenyl by Liu *et al.* [3], although the volume fraction accounting for the Meissner signal is still only 0.04%. In addition, a very similar gap size on K-dosed *p*-terphenyl has been observed with STM by Ren *et al.* [22], although they studied monolayer thin film grown *in-situ* instead of the single crystal we used in this study, and their spectral gap disappears at ~50 K. Although the STM gap shows no response to magnetic field up to 11T, the gap feature shows particle-hole symmetry, which is a signature for pairing gaps. A detailed discussion of the STM gaps is contained in supplementary information S3.

**Methods**

Single crystal samples of *p*-terphenyl were grown by the zone-refining method [23]. Their crystal structure was identified using X-ray diffraction [23]. Three crystals of similar size (around 0.2*0.5*0.5 mm) were used. Their surfaces were prepared through sublimation at <373 K for one hour in 1E-9 Torr vacuum. Photoemission measurements were carried out at the Stanford Synchrotron Radiation Lightsource (SSRL) beamline 5.4 with 32 eV linearly polarized light and 2E-11 Torr ultra-high vacuum. The experimental energy resolution was 14 meV. Fermi energy references were repeatedly obtained from the in-situ Au Fermi edge installed on the same sample manipulator. In-situ potassium dosing was performed using a commercial SAES getter source. It took the form of consecutive doses, with one dose corresponding to a heating current of 5.5 Amps that lasts 60 seconds. To avoid sample charging, significant potassium dosing was performed at 300 K before cooling to low temperature. The possibility of sample charging was ruled out through a comparison of spectral weight at different photon fluxes (Fig. S1), and the possibility of sample aging was ruled out by a comparison of spectral weight through various thermal and time cycles (Fig. S6). Other experiments to verify the general behavior and covering a wider range of photon energies were performed at the ALS beamline 4.0.1.

**Data availability**

The data that support the plots within this paper and other finding of this study are available from the corresponding author on reasonable request.

**Acknowledgements**

This work was funded by DOE project DE-FG02-03ER46066. We thank Drs. D. H. Lu, M. Hashimoto, and J. Denlinger for technical assistance on the ARPES measurements, and Sean Shaheen, Gang Cao, Gerald Arnold, Bruce Normand, Jennifer M. Reed, Justin Waugh and Miranda Thompson for help and valuable discussions. The photoemission experiments were performed at beamline 5-4 of the Stanford Synchrotron Radiation Lightsource, and beamlines 4.0.3 of the Advanced Light Source, Berkeley. The Stanford Synchrotron Radiation Lightsource is supported by the Director, Office of Science, Office of Basic Energy Sciences, of the U.S. Department of Energy under Contract No. DE-AC02-05CH11231. The Advanced Light Source is supported by the Director, Office of Science, Office of Basic Energy Sciences of the U.S. Department of Energy under Contract No. DE-AC02- 05CH11231.

**Author Contributions.**

E.C. prepared the single crystals of *p*-terphenyl. H.L., X.Z., S.P., T.N., J.G. and K.G. performed the experiment. H.L. and X.Z. analyzed the data. H.L., X.Z., and D.D. wrote the paper. D.D. conceived and directed the project. All authors read and commented on the paper.

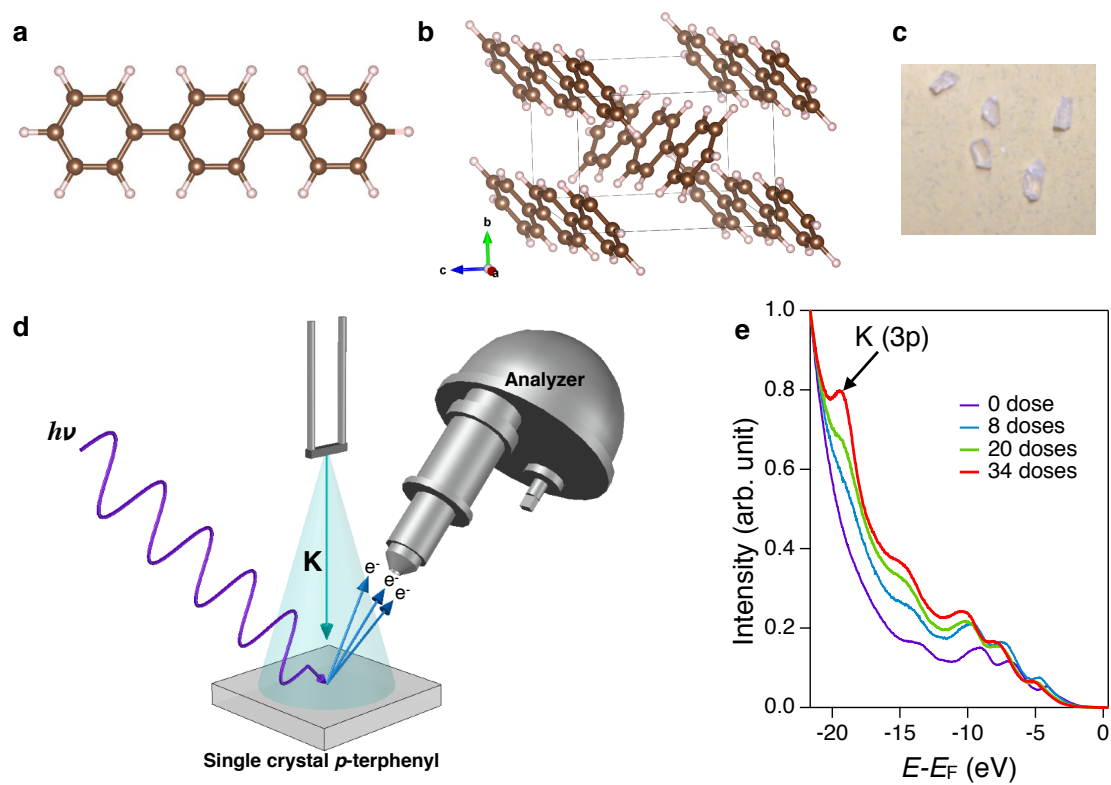

**Fig. 1, Materials, preparation, and broad overview. a**, *p*-terphenyl molecule. **b,** *p*-terphenyl molecules arranged in a single crystal. **c,** Pictures of our bulk crystals of *p*-terphenyl. **d**, The experimental schematic. K was repeatedly dosed onto the surfaces of the crystals, with photoemission spectra taken at these different doses. **e**, A wide overview of photoemission intensity vs. doping level. The growing intensity of the humps and peak around 18 eV (K 3p) in the photoemission spectra indicate the increasing doping level that follows the number of doses. The shift of the four spectral peaks of the valence band around 5, 7, 10, and 14eV shows the consistent change of chemical potential.

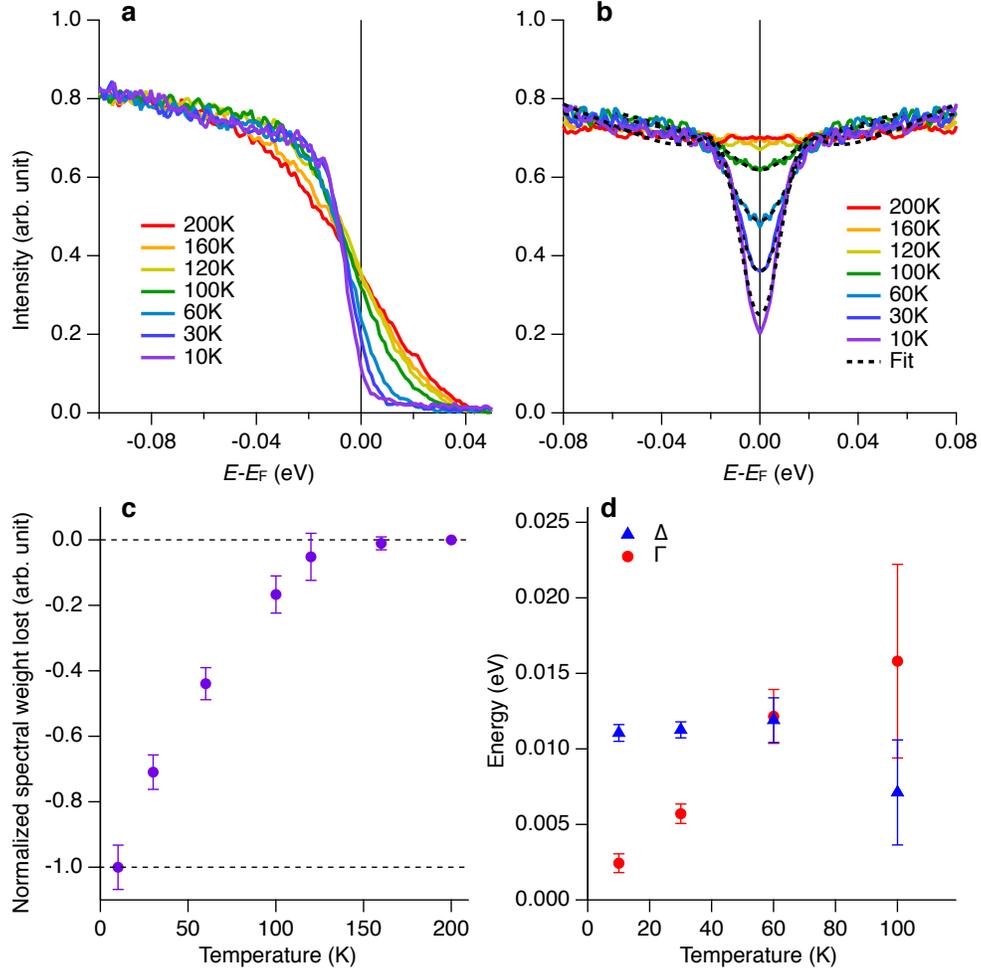

**Fig. 2, Spectral gaps from sample #3. a,** Temperature dependence of the very low energy photoemission spectra. **b,** The spectra of panel **a** symmetrized about $E_F$ so as to remove the effect of the Fermi function. This data clearly shows the presence of a gap at low temperatures, with the gap "filling in" as the temperature is increased. The dashed lines are fits to the data using equation 1, which has two key parameters – a gap $\Delta$ and a scattering rate $\Gamma$. **c,** Integrated spectral weight lost ($E_F \pm 3\text{meV}$) in the gap region vs. temperature, normalized to the lowest temperature measurement. The error bars denote the standard deviations of the spectral weight lost extracted using different energy ranges for normalizing the spectra. **d**, The gap $\Delta$ and scattering rate $\Gamma$ as a function of temperature from the fits of panel **b**. The error bar denotes the $3\sigma$ return from the fitting.

# Spectroscopic Evidence of Low Energy Gaps Persisting Towards 120 Kelvin in Surface-Doped *p*-Terphenyl Crystals

# Supplementary information

Haoxiang Li[1†*], Xiaoqing Zhou[1†*], Stephen Parham[1], Thomas Nummy[1], Justin Griffith[1], Kyle Gordon[1], Eric L. Chronister[2] and Daniel S. Dessau[1,3*]

[1]*Department of Physics, University of Colorado at Boulder, Boulder, CO 80309, USA*
[2]*Department of Chemistry, University of California, Riverside, CA, 92521, USA*
[3]*Center for Experiments on Quantum Materials, University of Colorado at Boulder, Boulder, CO 80309, USA*

† These authors contributed equally to this work

*Correspondence to: dan.dessau@colorado.edu
haoxiang.li@colorado.edu
xiaoqing.zhou@colorado.edu

**S1. Discussion of the weak spectral weight at $E_F$ in the doped compounds over a larger energy scale.**

With sufficient K-dosing, weak metallic spectral weight appeared near the chemical potential and the material became much more conductive. Compared to the spectral weight of the HOMO (Highest Occupied Molecular Orbital), the weight of these new doped features is ~100 times smaller (Fig. S2), i.e. we should not think of them as having rigidly doped the system into the LUMO (Lowest Unoccupied Molecular Orbital). This result has similarities to some other doped organic compounds such as picene and coronene, in which minimal electronic weight was found at the Fermi level [1,2], and is different from doped $C_{60}$ that does show strong spectral weight at $E_F$ [3].

In contrast to the depression of spectral weight over the first 12 meV or so that is discussed in detail in the main paper, this depletion of spectral weight covers a few hundred meV scale or more. A large energy scale reduction in spectral weight at $E_F$ is often termed a pseudogap, and may have relation to the large energy pseudogaps observed in cuprates [4,5,6], manganites [7], iridates [8], as well as the other organics (we note that some relatively low energy reductions of spectral weight, such as in underdoped cuprates, are also termed a pseudogap). Mechanisms to discuss such physics include strong electronic correlations (Mott physics) [9], polaronic effects [10], Coulomb gap effects [11], and doping heterogeneity [12], though it is too early to determine which, if any, of these is responsible for the very low near-$E_F$ spectral weight we observe in *p*-terphenyl. Note that these large energy pseudogaps may or may not be related to lower energy scale (10-50 meV) pseudogaps observed in some cuprates[13] and iridates[14], which have recently been discussed in terms of charge density waves [15] as well as of preformed Cooper pairs in the absence of long range phase coherence [16, 17].

On the other hand, we note that the extremely weak spectral weight near $E_F$ might be related to the extremely weak Meissner effect signal of the bulk material. It has been estimated from the Meissner signal strength that the shielding fraction of the bulk material is less than 0.05% in volume, the ratio of which seems to strongly depends on the growth/annealing condition. If this is the case, then a typical K-doped system might be regarded as a mixture of a dominant non-superconducting background which is not drastically different from the parent compound, and K-dosing-induced tiny superconducting grains embedded inside, which could correspond to strong ARPES spectra in the valence bands but very weak density of states near $E_F$ respectively. Such a scenario would also imply that a global zero resistivity state would be extremely difficult to achieve without significant improvement on the 0.04% level of volume ratio even with the presence of local pairing.

Beside the extremely inhomogeneous scenario as discussed above, there are other alternative scenarios that might address the unusually weak Meissner effect signal. The molecules in an organic solid are relatively far apart and not necessarily arrange for optimal wavefunction overlap, so coherent transport between molecules is in general difficult. Upon doping, a possible scenario is that the pairing emerges, but pairs are localized inside individual molecules, and have very little overlap with each other. As the pair size can be smaller than the separation between pairs, the system may be best described as in the "BEC limit" where the system remains a Bosonic insulator, and cannot efficiently shield the magnetic field nor provide strong near $E_F$ spectral weight. It is only with enhanced overlapping between wavefunctions that a crossover to the BCS limit might become possible. An example of such a scenario is proposed in reference [18], in which K-dosing drives a structural transition and enhances the overlap of π-orbitals, although it remains unclear how strong the overlap can be. In the cuprate case, there is ample experimental evidence of a very weak remnant diamagnetism signal above $T_C$ [19], which has been interpreted as originating from preformed Cooper pairs, which coincides with a finite spectral gap [16,17].

It is unclear whether the weak Meissner effect / weak spectral weight is dictated by the intrinsic physics of the system, or simply due to imperfection in the chemistry processes (to draw a parallel, early cuprate compounds only had 10% volume ratio [20]). The answer to this question is likely beyond the scope of a single experimental probe, and will require the combined efforts of the community.

**S2. The origin of the spectral gap**

As a spectroscopic probe, ARPES in general can map the characteristics of a gap well but cannot directly probe superconductivity itself. Here we give a more detailed discussion about why the pairing gap shows the best match to the phenomenology that we observed.

In general, if the gap we observe is due to a CDW (or SDW, Coulomb, polaronic effect), the existence of the weak Meissner effect would be an unusual coincidence, since a CDW (or SDW, Coulomb, polaronic effect) would have no direct connection to any Meissner effect. On the other hand, the presence of the weak Meissner effect with a similar temperature scale points towards the gap being related to pairing (but not necessarily long-range superconductivity). More importantly, the evolution of the in-gap spectral weight lost shown in Fig. 2c strongly resembles the one found in the pairing gap of cuprate high $T_C$ superconductor, where it was demonstrated to be a signature of the paring-gap phase (see Fig. 3 in ref [16]). In addition, in Fig 2d and Fig. S5, the fit parameters from the Dynes formula reveal a relatively small scattering rate ($\Gamma$) that is ~2meV for the low temperature (10 K) spectral weight. This value resembles the one from cuprates well below $T_C$ using the same Dynes formula, which accounts for the strong depletion of spectral weight inside the gap (shown in Fig. S7, also discussed in previous studies from both ARPES [17] and STM [21]). In contrast, the pseudogap (possibly from CDW [22] though the origin is still controversial) that is observed well above $T_C$ in the cuprates has a scattering rate that is typically a few tens of meV [17, 21], strongly filling the spectral gap (there is only a weak depletion of in-gap spectral weight). This indicates a clear distinction from the pairing gap well below $T_C$. On the other hand, a recent STM [23] work on K surface doped single layer p-terphenyl reveals a similar spectral gap that is symmetric above and below $E_F$ (a key feature of pairing gaps), and it showed no sign of a phase transition by charge ordering under varying temperature. In addition, spin density wave gaps found in iron-based superconductors also show a weak depletion in the spectrum that can be well distinguished from the low-energy pairing gap [24].

Moreover, our low-energy spectral gap is unlikely to be a Coulomb gap, which is a soft gap that usually develops over a large energy scale like 100meV or more [25,26]. A bipolaron at ~180meV has been proposed in the K doped terphenyl [27] from Raman spectroscopy. Again, this energy scale is way larger than our low-energy spectral gap. Based on the discussion above, we can conclude that the spectral gap that we found in the K doped terphenyl can to a high probability be accounted as a pairing gap instead of other origins.

**S3. Comparison to the STM study**

Ren *et al*. [23] recently presented a STM study on surface doping a single layer p-terphenyl deposited on an Au (111) surface. Compared with our study, the STM work gave a very similar gap size of 10-14 meV. As STM naturally probes both the occupied side and the un-occupied side, the data shows that the spectral gap is symmetric above and below $E_F$. Such symmetry is a key feature of the pairing gap. Their tomographic data also found no sign of a CDW transition under varying temperature.

On the other hand, the STM study found no signature of vortices in the presence of a magnetic field, and they found the gap to be insensitive to a magnetic field up to 11 Tesla. Both of these are at odds with the conventional wisdom of a superconductor. Nevertheless, such a behavior does not rule out a local pairing gap that precedes the superconducting state before the phase stiffness enables superconductivity. For instance, in the example of the Nernst effect in cuprate superconductors [28], it was believed that the prepairing gap does enable vortex-antivortex pairs. However, the vortices fluctuate at so fast a time scale that they cannot be detected by a static probe such as STM. In such a scenario, one might even argue that 11T is in fact a very small field compared to this scale of pairing temperature (the typical conversion is that 1T ~ 1K), and that the cuprates with similar temperature scales have critical fields over 100T. In fact, the author of ref [23] did discuss the local pairing gap as the most likely candidate for the origin.

We also note that there are certain differences between our own experiment that was carried out on a three-dimensional single crystal, and the STM study which was performed on a single layer *p*-terphenyl. To enable global superfluidity, generally the wavefunctions of preformed pairs need to overlap with each other, the process of which might well depend on the details of the structure and dimensionality. For instance, it was proposed that K-doping drives a 3D structural transition in the terphenyl crystal [18], which enhances the overlap between molecular orbitals and enables superconductivity. Such a transition is likely forbidden in a 2D case. It is also unclear whether the interface physics between the terphenyl and the Au substrate plays a role here. The STM gaps also show large zero-bias conductance even in the U-shape gap, which is different from typical superconductors.

## S4. Robustness and Reproducibility

To examined the robustness and reproducibility of our work, we performed the following operations:

As shown in Fig. S1, we varied the photon flux to check the potential surface charging. We found no signature of surface charging after K-dosing.

Fig. S3 exemplified the gap for different samples. While the statistics are different, the gap sizes are all comparable to each other.

Fig. S4 shows the comparison between a low temperature sample spectrum and the gold Fermi energy reference.

Fig. S5 shows data and fits similar to that of figure 2 of the main paper (sample 3) but from sample 2. This figure also shows the Dynes fit with and without the scattering term (panels b vs panel d). It is clear that even for a minimalist model the scattering term is needed.

Fig. S6 demonstrate the reproducibility before and after a temperature sweep from 10 K to 200 K, from sample 3. No sign of sample aging was found.

Fig. S7 demonstrate the phenomenological similarities between the cuprate superconductors and the p-terphenyl system. In particular, we stress that a lack of an apparent "coherence peak" is an intrinsic feature of the integrated spectral weight over momentum. As shown by the example of Fig. S7b, the integrated spectral weight often resembles a shifted Fermi edge even though the EDC at $k_F$ (Energy Distribution Curve) shows a clear "coherence peak". This can be understood as the EDC technique when measured at $k=k_F$ picks the sharpest feature of a whole spectrum. On the other hand, due to the lower statistics and the lack of a clear dispersion we cannot yet perform the same analysis on p-terphenyl. It is possible that with improved understanding of sample fabrication / doping / annealing it might be achieved in the future.

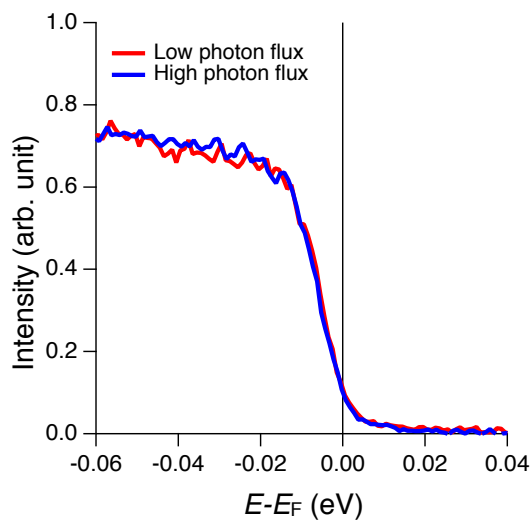

**Figure S1, Photoemission spectra with different photon flux.** The step edges of the two spectra with different photon flux are consistent. This indicates that there is no surface charging that could shift the spectral edge. The photon flux ratio is over 2.

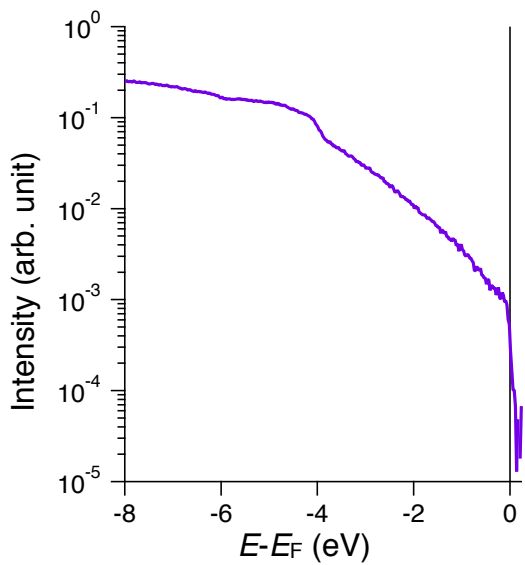

**Figure S2 Log scale plot of the spectrum over large energy range after effective surface doping with potassium.** The spectral peak around 5eV (also shown in Fig. 1e) is approximately 100 times larger than the spectral weight near $E_F$.

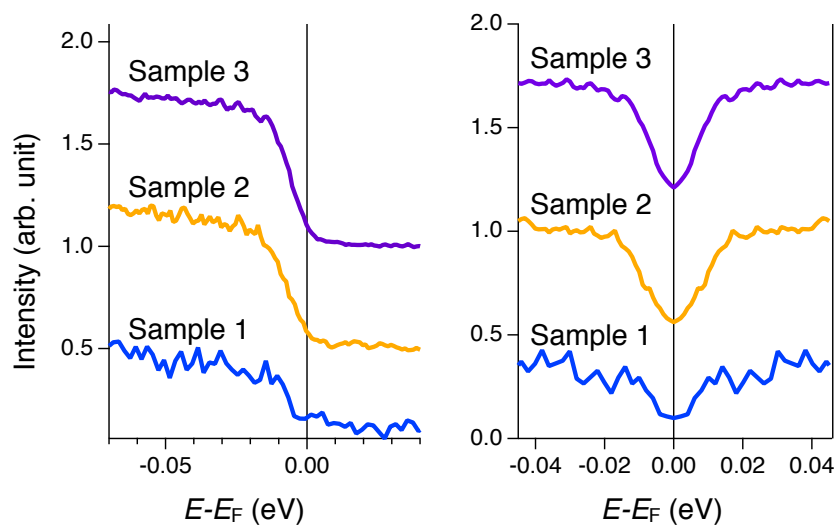

**Figure S3, Spectral gap at 10 K. a,** Photoemission spectra of three different samples with effective surface doping of potassium. All spectra show a clear spectral edge below the Fermi level. **b,** The corresponding symmetrized spectra that show the gap more clearly. Sample 2 is the sample from Fig. 1 and Fig. S5, while sample 3 is the sample from Fig. 2.

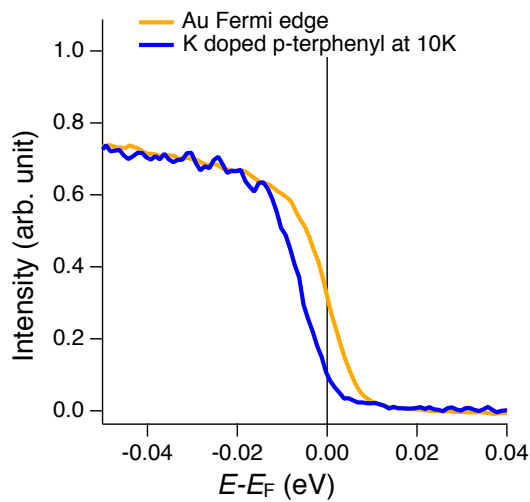

**Figure S4, Comparing sample spectrum to Fermi level reference.** The spectrum of the sample shows a leading edge well pushed away from that of the Au reference spectrum.

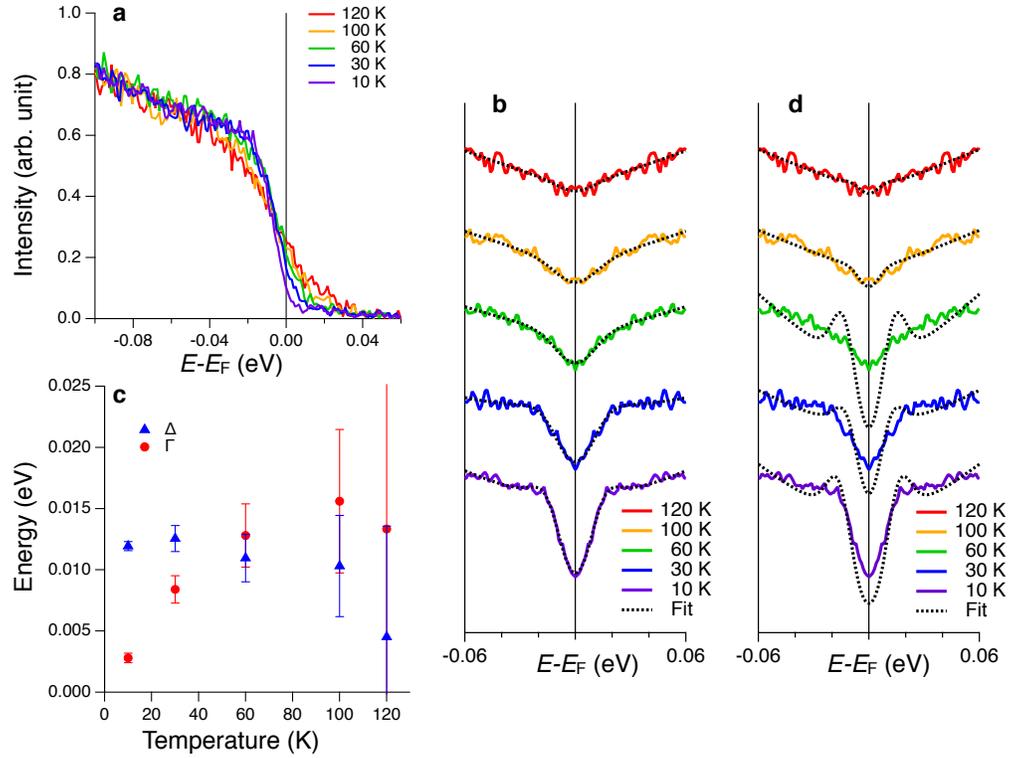

**Figure S5, Temperature evolution of the spectral gap in sample #2**. **a**, Temperature dependence of the photoemission spectra with effective doping of potassium. **b**, Same as (a) but symmetrized as well as fit to the Dynes formula. **c**, The superconducting gap (Δ) and the scattering rate (Γ) extracted from the fitting shown in panel **b**. **d**, Fitting without the scattering rate term (Γ). The coherence peak is much sharper when the scattering rate term is missing.

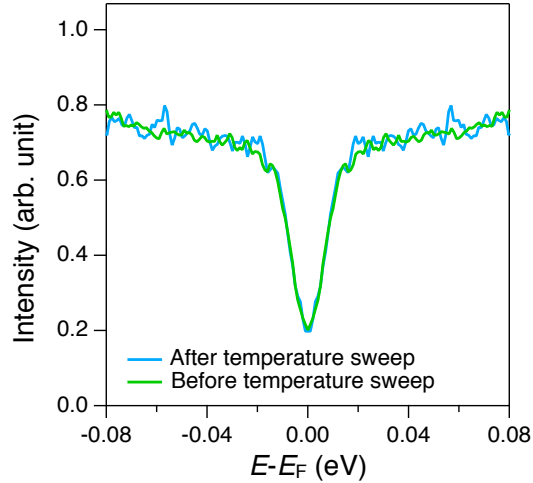

**Figure S6 Spectrum reproducibility after temperature sweep.** The two spectra of the sample show minimal changes before and after the temperature dependent measurement shown in Fig. 2.

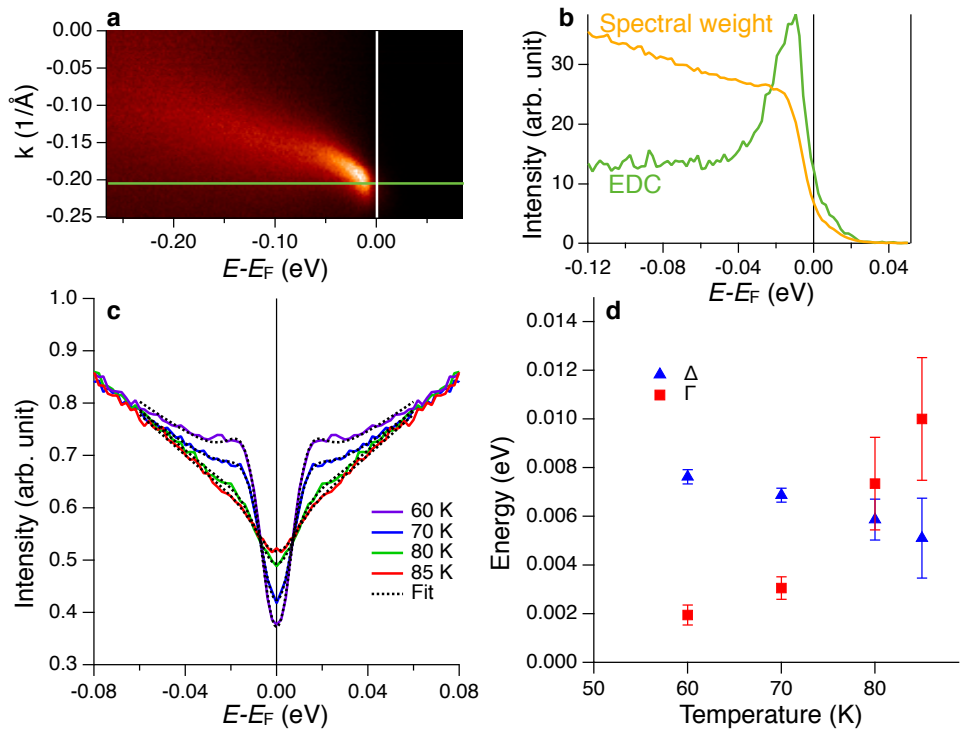

**Figure S7, Comparing to high Tc cuprate superconductor. a,** A typical ARPES spectrum of BSCCO superconductor taken at 60 K (Tc=85 K, lightly underdoped) in the near nodal region. **b,** The spectral weight (integrated over the momentum range of this cut) and an Energy Distribution Curve (EDC) at $k=k_F$ (green line in panel **a**). The EDC shows a strong coherence

peak, but not in the spectral weight. **c,** The symmetrized spectral weight at different temperatures and the corresponding fits to formula 1 of the main text. **d,** The extracted superconducting gaps ($\Delta$) and scattering rates ($\Gamma$), which have a temperature dependence similar to that of Fig 2c.